\documentclass[twocolumn,prl,showpacs,superscriptaddress]{revtex4}
\usepackage{amssymb}
\usepackage{amsmath}
\usepackage{graphicx}
\usepackage{epsfig}
\usepackage{amsfonts}

\begin{document}

\title{ Quantum State Reconstruction of Many Body System \\
Based on Complete Set of Quantum Correlations Reduced by Symmetry}
\author{X.F. Liu }
\email{liuxf@pku.edu.cn}
\affiliation{Department of Mathematics, Peking University, Beijing, 100871, China}
\author{C.P. Sun }
\email{suncp@itp.ac.cn}
\homepage{http://www.itp.ac.cn/~suncp}
\affiliation{Institute of Theoretical Physics, Chinese Academy of Sciences, Beijing,
100080, China}

\begin{abstract}
We propose and study a universal approach for the reconstruction of quantum
states of many body systems from symmetry analysis. The concept of minimal
complete set of quantum correlation functions (MCSQCF) is introduced to
describe the state reconstruction. As an experimentally feasible physical
object, the MCSQCF is mathematically defined through the minimal complete
subspace of observables determined by the symmetry of quantum states under
consideration. An example with broken symmetry is analyzed in detail to
illustrate the idea.
\end{abstract}

\maketitle

\emph{Introduction:} The concept of quantum state is fundamental in quantum
physics. Quantum state can provide us with a complete knowledge of the
considered system to predict the possible results of any measurement on the
system \cite{QM}. But the puzzling property of quantum coherence, which is
intrinsic to quantum state, imposes insurmountable limitations on our
ability to fully reconstruct quantum state of a single quantum system by
devising a complete set of measurements on the system \cite{IM}.

Actually, we cannot make a measurement on a single quantum system without
back-action and the non-cloning theorem \cite{NC} forbids us from producing
an exact copy of some quantum system in an unknown state. But on the other
hand, if identical copies of quantum system, or an ensemble, is initially
prepared in the same quantum state, it is possible to estimate the unknown
state of the quantum system by carrying out appropriate measurements on each
copy \cite{fano}. This is just the idea of quantum tomography, which has
recently become a fashion in quantum information physics, motivating many
theoretical and experimental investigations \cite{EXP}.

We notice that most works on quantum tomography or other quantum state
reconstruction methods focus on continuous variable system and single qubit
system. In this letter we will deal with the many particle case, which
possesses distinguished features such as quantum entanglements \cite%
{Wootters:98}. It has become well known that quantum correlation and the
corresponding Green function are important conceptual tools to probe the
non-local nature of many body system. On the other hand, in the study of
quantum information, quantum entanglement turns out to be a crucial concept
and thus becomes a fundamental element in the research about the
non-locality of quantum system. Hence, it is doubtlessly important and
meaningful to probe the possible connection between quantum entanglement and
quantum correlation \cite{ring}. Along this line, some progress has been
made. For example, for the spin chain system with the symmetry of $SO(2)$, a
connection has been revealed between the concurrence, which is a measure of
quantum entanglement, and the correlation functions of first order and
second order, which is supposed to reflect the property of correlation \cite%
{Wang}. This connection suggests a possible connection between quantum
entanglement and quantum phase transition \cite{Kit,lin, Qian} and prompts
us to introduce in this letter a general framework for studying such
connection.

Our starting point is the basic observation that complete information about
quantum state of many body system can be obtained from the expectation
values of properly chosen single body observables and their correlation
functions of various orders \cite{Yang}. And our idea is to reduce the
number of times of measurement needed to determine a quantum state by making
full use of the symmetry that the state is supposed to possess. The main
purpose of this letter is to make clear the mechanism of this reduction and
present interesting examples to illustrate it. To this end we will introduce
the minimal complete subspace of observables (MCSO) associated with a
symmetry group, which mathematically determines the quantum state with this
symmetry, and the minimal complete set of quantum correlation functions
(MCSQCF) associated with the MCSO, which determines the quantum state
physically in some sense.

\emph{State Reconstruction Based on Correlation Functions:} Throughout this
letter, let $V=\otimes_{j=1}^{m}W_{j}\triangleq W_1\otimes
W_2\otimes\cdots\otimes W_m$ be the Hilbert space of a many spin system.
Here the $W_{i}$ 's stand for the state space $W$ of a spin system. For an
arbitrary finite dimensional Hilbert space $S$ we denote by ${\mathcal{A}}%
(S) $ the set of Hermitian operators on $S$. It is well known that ${%
\mathcal{A}}(S)$ is an Euclidean space with the inner product $\langle \ ,\
\rangle $ defined as $\langle A,B\rangle =\mathrm{tr}(AB),\;\forall \ A,B\in
{\mathcal{A}}(S)$, where $\mathrm{tr}$ is the trace over $W$.

Now let $\rho \in {\mathcal{A}}(V)$ be a density operator and $\{A_{i}\}$ a
basis of ${\mathcal{A}}(V)$. Then $\rho $ can be written as a linear
combination of $A_{i}$'s: $\rho =\sum_{i}c_{i}A_{i}$. We notice that the
coefficients $c_{i}$'s are determined by the expectation values $\{\mathrm{%
tr(}\rho A_{i})\}$ of the observable $A_{i}$ in the state $\rho $. In
particular, if $\{A_{i}\}$ is an orthonormal basis then we have $c_{i}=%
\mathrm{tr}\rho A_{i}$. Obviously, from mathematical point of view it is
trivially true that $\rho $ is completely determined by $\{\langle \rho
,A_{i}\rangle \}$. But physically it means the possibility of determining a
quantum state by measuring some properly chosen observables.

To be precise, let us choose a basis $\{O_{i}\}$ of ${\mathcal{A}}(W)$. The $%
O_{i}$'s can be understood as observables of single particle system. Then $%
\{\otimes _{k=1}^{m}O_{i_{k}}\}$ is a basis of ${\mathcal{A}}(V)$, where
each $O_{i_{j}}$ belongs to $\{O_{i}\}$. Hence we can identify the state $%
\rho $ by the quantities
\begin{equation}
\mathrm{tr(}\rho \,\otimes _{k=1}^{m}O_{i_{k}})\triangleq \langle \otimes
_{k=1}^{m}O_{i_{k}}\rangle _{\rho },
\end{equation}%
which are just correlation functions of the considered system.

For spin $1/2$ system interesting things happen\cite{ring,Wang}. In this
case we can take $\{O_{i}\}$ to be $\{\sigma _{0},\sigma _{1},\sigma
_{2},\sigma _{3}\}$, where $\sigma _{0}$ is the identity operator and $%
\sigma _{1},\sigma _{2},\sigma _{3}$ are the Pauli operators $\sigma
_{x},\sigma _{y},\sigma _{z}$. Let us introduce the standard correlation
functions $G_{i_{1}i_{2}\cdots i_{n}}^{\rho }\triangleq \langle \otimes
_{k=1}^{n}\sigma _{i_{k}}\rangle _{\rho },$ associated with the state $\rho $%
. Clearly, from both theoretical and experimental points of view standard
correlation functions are accessible physical quantities. According to the
discussion above, for spin $1/2$ system a state is determined by these ideal
quantities.

We are not so lucky when we consider higher spin system. The main reason is,
when the dimension of $W$ is larger than $2$ it is impossible to span ${%
\mathcal{A}}(W)$ with just the spin operators and identity operators. This
forces us to consider the correlation functions not related to the spin
operators in a direct way. From experimental point of view such correlation
functions are less physical than the standard correlation functions. But
from theoretical point of view there is no essential difference between them.

Now a natural question is: how many correlation functions are needed to
determine a quantum state? Obviously, if we have no knowledge about the
state, then the number of correlation functions needed is exactly the
dimension of ${\mathcal{A}}(V)$. But if we are sure that the state possesses
some symmetry, then it turns out that the number might be greatly reduced.
Let us elucidate this point in detail as follows.

\emph{Reduction to the Minimal Complete Subspace of Observables by Symmetry: }
Now,we discuss symmetry of density operator and the role symmetry can play in
the reconstruction of density operator.

The symmetry group of the density operator $\rho \in {\mathcal{A}}(V)$ is
defined to be the subgroup $U_{\rho }$ of $U(n)$ with the property $u\rho
u^{\dag }=\rho ,\;\forall \ u\in U_{\rho }$. If $U$ is a subgroup of $%
U_{\rho }$ we will say that $\rho $ possesses the symmetry of $U$. In
practice, it is highly possible that only partial information about the
symmetry of a density operator is available. In other words, one may only be
sure that some group is a subgroup of the symmetry group of a density
operator. This being the case, it is desirable to know how to identify a
density operator among a family of density operators whose symmetry groups
contain a specific unitary group $U$ (a subgroup of $U(n)$ by definition).
This is the idea underlying the following discussion.

A symmetry of $U$ determines a family $O_{U}$ of observables in the
following way: $A\in O_{U}$ if and only if $uAu^{\dag }=A,\;\forall \ u\in U$%
. By definition, if $A\in O_{U}$ then its symmetry group contains $U$ as a
subgroup and vice versa. Thus $\rho $ belongs to $O_{U}$ if and only if it
possesses the symmetry of $U$. Clearly, $O_{U}$ is a subspace of ${\mathcal{A%
}}(V)$. We call it the MCSO associated with $U$.

As ${\mathcal{A}}(V)$ is an inner product space it can be decomposed as ${%
\mathcal{A}}(V)=O_U\oplus O_U^{\bot},$ where $O_U^{\bot}$ is the orthonormal
complement to $O_U$. Let $\{A_i\}$ be an arbitrary basis of $O_U$. Then each
element $A$ of $O_U$ is uniquely determined by the quantities $\{\langle
A,A_i\rangle\}$. This justifies calling $O_U$ complete. Consequently, if $%
\rho$ possesses the symmetry of $U$ then the number of measurements needed
to determine $\rho$ is just the dimension of the MCSO associated with $U$.

Next let us determine the structure of $O_{U}$.

\textbf{Proposition 1.}\textit{\ If }$U$\textit{\ is connected, then }$%
\,iO_{U}$\textit{\ is the centralizer of the Lie algebra of }$U$\textit{\ in
}$u(n)$\textit{, the Lie algebra of }$U(n)$\textit{, where }$n$\textit{\ is
the dimension of }$V$. ( Here $i=\sqrt{-1}$.)

Proof. We notice that $i\,{\mathcal{A}}(V)$ is exactly the Lie algebra of $%
U(n)$. Let $i\,A$ be an arbitrary element of the Lie algebra of $U$ and $i\,B
$ an arbitrary element of $i\,O_{U}$. Then we have $iB=\exp ({it\,A}%
)(i\,B)\exp (-{it\,A})${\ or}%
\begin{equation}
iB=(\mathrm{Ad}{e}^{{it\,A}})(iB)=e^{\,it\,\mathrm{ad}\,A}(i\,B).
\end{equation}%
It follows that $\,iO_{U}$ is a subset of the centralizer of the Lie algebra
of $U$. That they are actually identical is now a consequence of the fact
that if $U$ is connected, then it can be generated by $1$-parameter
subgroups. This completes the proof of the proposition.

When $U$ is a $1$-parameter subgroup the structure of $O_{U}$ can be given a
concise description. Let us consider this case in more detail. Suppose that $%
U=\exp {(-itA)}$ where $A\in {\mathcal{A}}(V)$, then according to
Proposition 1, an element $B$ belongs to $O_{U}$ if and only if $[A,B]=0$.
As $A$ is an Hermitian operator there exists an orthonormal basis $\gamma $
of $V$ with respect to which it has the matrix representation of block
diagonal form (the meaning of notation here should be self-evident): $A=%
\mathrm{diag}\{A_{1},A_{2},\cdots ,A_{n}\},$ where $A_{i}=\lambda
_{i}I_{n_{i}}$, $\lambda _{i}$ being an eigenvalue of $A$ and $I_{n_i}$ is
the unit matrix of rank $n_{i}$, $n_{i}$ being the multiplicity of $\lambda
_{i}$. From this observation, the following proposition follows directly.

\textbf{Proposition 2}\textit{. If }$U$\textit{\ is connected then an
element }$B$\textit{\ lies in }$O_{U}$\textit{\ if and only if its matrix
representation takes the form }$B=diag\{B_{1},B_{2},\cdots ,B_{n}\},$\textit{%
\ with respect to the same basis }$\gamma $\textit{\ defined above, where }$%
B_{i}$\textit{\ is an }$n_{i}$\textit{\ by }$n_{i}$\textit{\ matrix.}

\emph{Symmetry of Reduced Density Operators:} In this section we consider
symmetry of the so called reduced density operator, which is derived from
the density operator after a process of taking partial trace. We are
interested in the symmetry that survives such process.

Let $V=V_{1}\otimes V_{2}$ and $\rho $ a density operator in ${\mathcal{A}}%
(V)$. The reduced density operator $\mathrm{tr}_{2}\rho $ is defined to be
an element of ${\mathcal{A}}(V_{1})$ satisfying the relations $(x_{1},(%
\mathrm{tr}_{2}\rho )x_{2})=\sum_{i}(x_{1}\otimes y_{i},\rho \,x_{2}\otimes
y_{i}),\forall \ x_{1},x_{2}\in V_{1},$ where $\{y_{i}\}$ is a basis of $%
V_{2}$. Note that we have used the same notation $(\ ,\ )$ for the inner
products in the different spaces. It is easily check that $\mathrm{tr}%
_{2}\rho $ is also a density operator. Now the following lemma comes as a
direct consequence of the definitions. We would rather omit the proof.

\textbf{Lemma 1}\textit{. If }$\rho $\textit{\ possesses the symmetry of }$%
U_{1}\otimes U_{2}$\textit{\ then }$tr_{2}\rho $\textit{\ possesses the
symmetry of }$U_{1}$\textit{.}

Although the idea of studying how symmetry survives the reduction process
might seem trivial, there are nontrivial examples to support it. Let us
consider two models of such examples. For convenience, from now on we will
use the Dirac notation.

Suppose that $A\in {\mathcal{A}}(V)$ has the form $A=A_{1}\otimes 1+1\otimes
A_{2}$, where $A_{1}\in {\mathcal{A}}(V_{1})$ and $A_{2}\in {\mathcal{A}}%
(V_{2})$. Let $|\psi \rangle $ be a normalized eigenvector of $A$. Then it
is easily seen that the density operator $\rho =|\psi \rangle \langle \psi |$
possesses the symmetry of the $1$-parameter subgroup $\exp {(it\,A)}$ that
can be written as $\exp {(it\,A)}=\exp {(it\,A_{1})}\otimes \exp {(it\,A_{2})%
}.$ Thus the reduced density operator $\mathrm{tr}_{2}\rho $ possesses the
symmetry of $\exp {(it\,A_{1})}$ according to Lemma~1.

In the special case that $\rho$ has the symmetry of $1$-parameter subgroup
there is another mechanism of passing symmetry to the reduced density
operator, which is different from the mechanism shown above. Let $A$ be an
Hermitian operator of the form $A=A_1\otimes A_2$ where $A_1\in {\mathcal{A}}%
(V_1)$ and $A_2\in {\mathcal{A}}(V_2)$ and $|\psi\rangle$ a normalized
eigenvector of $A$. Then for the density operator $\rho=|\psi\rangle\langle%
\psi|$ we have the following result.

\textbf{Lemma 2. }\emph{The reduced density operator }$tr_{2}\rho $\emph{\
possesses the symmetry of }$\exp (it\,A_{1})$\emph{\ if the eigenvalue
corresponding to }$|\psi \rangle $\emph{\ is nonzero.}

Proof. Let $\{|\phi _{i}\rangle \}$ be an orthonormal basis of $V_{1}$
consisting of the eigenvectors of $A_{1}$ and $\{|\varphi _{j}\rangle \}$ an
orthonormal basis of $V_{2}$ consisting of the eigenvectors of $A_{2}$.
Suppose that $A_{1}|\phi _{i}\rangle =\lambda _{i}|\phi _{i}\rangle $ and $%
A_{2}|\varphi _{j}\rangle =\mu _{j}|\varphi _{j}\rangle .$ Since $|\psi
\rangle $ is an eigenvector of $A$ there exist $\lambda _{k}$ and $\mu _{l}$
such that $A|\psi \rangle =\lambda _{k}\mu _{l}|\psi \rangle .$ By
definition, for arbitrary two elements $|\phi _{i_{1}}\rangle $ and $|\phi
_{i_{2}}\rangle $ of the basis $\{|\phi _{i}\rangle \}$ we have $\langle
\phi _{i_{1}}|\mathrm{tr}_{2}\rho |\phi _{i_{2}}\rangle =\sum_{j}\langle
\varphi _{j},\phi _{i_{1}}|\psi \rangle \langle \psi |\phi _{i_{2}},\varphi
_{j}\rangle ,$ where $|\phi _{i_{2}},\varphi _{j}\rangle =|\phi
_{i_{2}}\rangle \otimes |\varphi _{j}\rangle $. It follows that $\langle
\phi _{i_{1}}|\mathrm{tr}_{2}\rho |\phi _{i_{2}}\rangle =0,$ unless $\lambda
_{i_{1}}=\lambda _{i_{2}}$. On the other hand, we have
\begin{equation}
P=e^{it(\lambda _{i_{1}}-\lambda _{i_{2}})}\langle \phi _{i_{1}}|\,\mathrm{tr%
}_{2}\rho \,|\phi _{i_{2}}\rangle .
\end{equation}%
for $P\triangleq \langle \phi _{i_{1}}|\exp ({it\,A_{1})}\mathrm{tr}_{2}\rho
\exp (-{it\,A_{1})}|\phi _{i_{2}}\rangle .$This just means that
\begin{equation}
P=\left\{
\begin{array}{cc}
\langle \phi _{i_{1}}|\,\mathrm{tr}_{2}\rho \,|\phi _{i_{2}}\rangle , &
\mathrm{if}\;\lambda _{i_{1}}=\lambda _{i_{2}} \\
0, & \mathrm{otherwise}.%
\end{array}%
\right.
\end{equation}%
Thus we conclude that $\exp {(it\,A_{1})}\,\mathrm{tr}_{2}\rho \,\exp {%
(-it\,A_{1})}=\mathrm{tr}_{2}\rho $ and complete the proof.

\emph{Minimal Complete Set of Correlation Functions: }In this section the
problem of reconstruction of quantum state is reduced to the problem of
determination of a set of correlation functions.

Let $\{O_{i}\}$ be a basis of ${\mathcal{A}}(W)$. Then $\{\otimes
_{k=1}^{m}O_{i_{k}}\}$ is a basis of ${\mathcal{A}}(V)$. We call it the
measurement basis of ${\mathcal{A}}(V)$ associated with $\{O_{i}\}$. A
subset of the measurement basis of ${\mathcal{A}}(V)$ is called a
reconstruction basis of the MCSO associated with $\{O_{i}\}$ if each element
of the MCSO can be written as a linear combination of elements in the subset
and a reconstruction basis is called minimal if it does not contain any
smaller reconstruction basis. Note that according to this definition
elements of a reconstruction basis of the MCSO do not necessarily belong to
the MCSO.

\textbf{Proposition 3.}\textit{\ There exists a unique minimal
reconstruction basis of an MCSO associated with each }$\{O_{i}\}$\textit{.}

Proof. Choose an arbitrary basis $\beta $ of the considered MCSO and define
a subset $\gamma _{1}$ of the measurement basis $\{\otimes
_{k=1}^{m}O_{i_{k}}\}$ as follows: an element lies in $\gamma _{1}$ if and
only if it appears as a non zero term in the expression of some element of $%
\beta $ in terms of $\{\otimes _{k=1}^{m}O_{i_{k}}\}$. Clearly $\gamma _{1}$
is a minimal reconstruction basis associated with $\{O_{i}\}$. We claim that
it is in fact the unique minimal reconstruction basis of the considered MCSO
associated with $\{O_{i}\}$.

Let $\gamma _{2}$ be another minimal reconstruction basis associated with $%
\{O_{i}\}$. Then each element of $\beta $ can be expressed as a linear
combination of the elements of $\gamma _{2}$ or as a linear combination of
the elements of $\gamma _{1}$. Thus by equating the two expressions we
conclude that $\gamma _{1}\subseteq \gamma _{2}$, considering the fact that
both $\gamma _{1}$ and $\gamma _{2}$ are subsets of the measurement basis $%
\{\otimes _{k=1}^{m}O_{i_{k}}\}$. That $\gamma _{1}$ and $\gamma _{2}$ are
actually identical now follows from the minimality of $\gamma _{2}$. This
completes the proof of the proposition.

We are now in a position to introduce the concept of MCSQCF. Let $O_{U}$ be
an MCSO and $\{\otimes _{k=1}^{m}O_{j_{k}}\}$ the minimal reconstruction
basis of $O_{U}$ associated with $\{O_{i}\}$. Please note the notational
difference between $\{\otimes _{k=1}^{m}O_{j_{k}}\}$and $\{\otimes
_{k=1}^{m}O_{i_{k}}\}$. Here the former is used to denote a subset of the
latter. With respect to $\{\otimes _{k=1}^{m}O_{i_{k}}\}$, the MCSQCF
associated with $O_{U}$ is defined to be the set $\{\langle \otimes
_{k=1}^{m}O_{j_{k}}\rangle _{\rho }\}$ of quantum correlation functions
where $\rho $ stands for a density operator in $O_{U}$. The next proposition
shows that each $\rho $ in $O_{U}$ is completely determined by $%
\{\langle\otimes _{k=1}^{m}O_{j_{k}}\rangle _{\rho }\}$ but not by any
proper subset of it. This is just what \textquotedblleft complete" and
\textquotedblleft minimal" mean in the definition of MCSQCF.

\textbf{Proposition 4.}\textit{\ A density operator }$\rho $\textit{\ in }$%
O_{U}$\textit{\ can be expressed in terms of the MCSQCF }$\{\langle \otimes
_{k=1}^{m}O_{j_{k}}\rangle _{\rho }\}$\textit{, but cannot be expressed in
terms of any proper subset of }$\{\langle \otimes _{k=1}^{m}O_{j_{k}}\rangle
_{\rho }\}$\textit{, regardless of the basis of }$V$\textit{\ on which }$%
\rho $\textit{\ acts.}

Proof. Let $\{B_{i}\}$ be an orthonormal basis of $O_{U}$. We then have $%
\rho =\sum_{i}\langle \rho ,B_{i}\rangle B_{i}=\sum_{i}(\mathrm{tr}\rho
B_{i})B_{i}$, where each coefficient $\mathrm{tr}\rho B_{i}$ can be written
as a linear combination of $\{\langle \otimes _{k=1}^{m}O_{j_{k}}\rangle
_{\rho }\}$. Moreover, each $\langle \otimes _{k=1}^{m}O_{j_{k}}\rangle
_{\rho }$ will appear in some $\mathrm{tr}\rho B_{i}$ by the minimality of $%
\{\otimes _{k=1}^{m}O_{j_{k}}\}$. Hence $\rho $ can be written as
\begin{equation}
\rho =\sum_{j_{1}j_{2}\cdots j_{m}}\langle \otimes
_{k=1}^{m}O_{j_{k}}\rangle _{\rho }B_{j_{1}j_{2}\cdots j_{m}},
\end{equation}%
where $B_{j_{1}j_{2}\cdots j_{m}}$ is a linear combination of $\{B_{i}\}$,
which cannot be a zero matrix with respect to any basis of $V$ since $%
\{B_{i}\}$ is linearly independent. This proves the proposition.

\emph{Example with Broken Symmetry : }\ Now let us present a concrete
example to illustrate the results obtained above. We consider the case that $%
W$ is the state space of spin $1/2$ particle. Let $\ |\uparrow \rangle $ and
$|\downarrow \rangle $ be the normalized eigenvectors of the Pauli operators
$\sigma _{z}$ and take $V=V_{1}\otimes V_{2},V_{1}=W_{1}\otimes W_{2},$ $%
V_{2}=W_{3}\otimes \cdots \otimes W_{m}$.

We suppose the state $\rho $ to be measured possess the symmetry generated
by $A=\otimes _{j=1}^{m}\sigma _{z}^j$, namely the symmetry of $\exp {(it\,A)%
}$, where $m$ is even. Then according to Lemma~2, the reduced density
operator $\tilde{\rho}\triangleq\mathrm{tr}_{2}\rho $ possesses the symmetry
of $U=\exp {(it\,A_{1})}$ with $A_{1}=\sigma _{z}\otimes \sigma _{z}$.
Consequently, by Proposition~2, with respect to the basis $\{|\uparrow
\uparrow \rangle ,|\downarrow \downarrow \rangle ,|\uparrow \downarrow
\rangle ,|\downarrow \uparrow \rangle \}$, the reduced density operator is
of the 8-vertex form
\begin{equation}
\tilde{\rho}=\left(
\begin{array}{cccc}
r_{1} & z_{1} & \  & \  \\
z_{1}^{\ast } & r_{2} & \  & \  \\
\  & \  & r_{3} & z_{2} \\
\  & \  & z_{2}^{\ast } & r_{4}%
\end{array}%
\right) ,
\end{equation}%
where the $r_i$'s are real numbers with the restriction $\sum_ir_i=1$ and $%
z_i^*$ stands for the complex conjugation of $z_i$. One can check that $%
\{1\otimes 1,\ 1\otimes \sigma _{z},\ \sigma _{z}\otimes 1,\ \sigma
_{x}\otimes \sigma _{x},\ \sigma _{y}\otimes \sigma _{y},\ \sigma
_{z}\otimes \sigma _{z},\ \sigma _{x}\otimes \sigma _{y},\ \sigma
_{y}\otimes \sigma _{x}\}$ is the unique reconstruction basis of $O_{U}$
associated with $\{\sigma _{i}\}$ and that $\{G_{00}^{\tilde{\rho}},\
G_{03}^{\tilde{\rho}},\ G_{30}^{\tilde{\rho}},\ G_{11}^{\tilde{\rho}},\
G_{22}^{\tilde{\rho}},\ G_{33}^{\tilde{\rho}},\ G_{12}^{\tilde{\rho}},\
G_{21}^{\tilde{\rho}}\}$ is the cprresponding MCSQCF. The explicit
expression of $\tilde{\rho}$ in terms of the MCSQCF is as follows: $%
r_{1}=(G_{33}^{\tilde{\rho}}+G_{30}^{\tilde{\rho}}+G_{03}^{\tilde{\rho}%
}+1)/4,r_{2}=(G_{33}^{\tilde{\rho}}-G_{30}^{\tilde{\rho}}-G_{03}^{\tilde{\rho%
}}+1)/4,r_{3}=(G_{30}^{\tilde{\rho}}-G_{33}^{\tilde{\rho}}-G_{03}^{\tilde{%
\rho}}+1)/4,\quad $\ $r_{4}=(G_{03}^{\tilde{\rho}}-G_{33}^{\tilde{\rho}%
}-G_{30}^{\tilde{\rho}}+1)/4,z_{1}=(G_{11}^{\tilde{\rho}}-G_{22}^{\tilde{\rho%
}}+iG_{12}^{\tilde{\rho}}+iG_{21}^{\tilde{\rho}})/4,z_{2}=(G_{11}^{\tilde{%
\rho}}+G_{22}^{\tilde{\rho}}+iG_{21}^{\tilde{\rho}}-iG_{12}^{\tilde{\rho}%
})/4.$

This example has a non-trivial connection with the Ising model with the
Hamiltonian
\begin{equation}
H=-J\sum_{j=1}^{m-1}\sigma _{z}^{j}\sigma _{z}^{j+1}+g\sum_{j=1}^{m}\sigma
_{z}^{j},\;J>0.
\end{equation}%
When the external field $g=0$, this model possesses the $SO(2)\otimes Z_{2%
\text{ }}$symmetry where $Z_{2\text{ }}$represents the rotations of all spins
by 180 degrees about the x axis while $SO(2)$ represents the rotations of all
spins by arbitrary degrees about the z axis. However, this system has two
degenerate ferromagnetically ordered ground states $|\uparrow \rangle ^{\otimes
m}$ and $|\downarrow \rangle ^{\otimes m}$, which break the discrete symmetry
$Z_{2}$. Obviously, as a superposition of these two ground states, the cat
state$|\psi \rangle =\alpha |\uparrow \rangle ^{\otimes m}+\beta |\downarrow
\rangle ^{\otimes m}$ possess the symmetry of $\exp {(it\,A)}$ mentioned above
since it is the eigenvector of $A=\otimes _{j=1}^{m}\sigma _{z}^{j}$
corresponding to the eigenvalue $1$, and then the unknown $\alpha $ and $\beta
$ satisfying $|\alpha |^{2}+|\beta |^{2}=1$ can be determined by the above
quantum state reconstruction method. When $g\neq 0 $, the reduced density
operator $\tilde{\rho}=$ $\mathrm{tr}_{2}\rho $ derived from the ground state
or the thermal equilibrium state(a function of hamiltonian)is of the 6-vertex
form for the existence of the total spin projection $\sum_{j}\sigma _{z}^{j}.$
This result has appeared in many places, e.g., in the references
\cite{ring,Wang}

\emph{Concluding remarks :} Based on the viewpoint of symmetry, we have
proposed a quantum state reconstruction method for density operators or
reduced density operators of the many body system. Our central idea is to
utilize the MCSQCF determined by the symmetry of quantum states rather than
the symmetry of the Hamiltonian. The present work might be regarded as a
generalization of the ``particle creation-annihilation expression" for the
reduced density matrices of identical particles in Ref.\cite{Yang}, which
reveals the essence of quantum condensation. Thus, we hope that the results
in this letter can be directly applied to explore the quantum and classical
critical phenomenon for various quantum many body systems.

This work is supported by the NSFC with grant Nos. 90203018, 10474104 and
60433050. It is also funded by the National Fundamental Research Program of
China with Nos. 2001CB309310 and 2005CB724508.

\end{document}